\documentclass[aps,prb,twocolumn,showpacs,showkeys,floatfix]{revtex4}

\usepackage{graphicx}
\graphicspath{{figs/}}
\begin{document}
\title{Manipulation of heat current by the interface between graphene and white graphene}
\author{Jin-Wu~Jiang}
    \altaffiliation{Electronic address: phyjj@nus.edu.sg}
    \affiliation{Department of Physics and Centre for Computational Science and Engineering,
             National University of Singapore, Singapore 117542, Republic of Singapore }
\author{Jian-Sheng~Wang}
    \affiliation{Department of Physics and Centre for Computational Science and Engineering,
                 National University of Singapore, Singapore 117542, Republic of Singapore }

\date{\today}
\begin{abstract}
We investigate the heat current flowing across the interface between graphene and hexagonal boron nitride (so-called white graphene) using both molecular dynamics simulation and nonequilibrium Green's function approaches. These two distinct methods discover the same phenomena that the heat current is reduced linearly with increasing number of carbon atom at the interface, and the zigzag interface causes stronger reduction of heat current than the armchair interface. These phenomena are interpreted by both the lattice dynamics analysis and the transmission function explanation, which both reveal that the localized phonon modes at interfaces are responsible for the heat management. The room temperature interface thermal resistance is about $7\times10^{-10}$m$^{2}$K/W in zigzag interface and $3.5\times10^{-10}$m$^{2}$K/W in armchair interface, which directly results in stronger heat reduction in zigzag interface. Our theoretical results provide a specific route for experimentalists to control the heat transport in the graphene and hexagonal boron nitride compound through shaping the interface between these two materials.
\end{abstract}

\maketitle

\pagebreak

\section{introduction}
The one-atom-thick graphene sheet possesses unique electronic properties based on the zero energy gap in its electron band structure. However, a finite band gap is of crucial importance for applications in field-effect transistor devices. Therefore, a meaningful work is to realize band gap opening in the electron band structure of graphene. Both experimental\cite{CiL} and theoretical\cite{FanX,LiuY} groups have recently demonstrated a novel approach to accomplish tunable band gap through growing hexagonal boron nitride islands within graphene sheet. This improvement benefits the practical development of graphene-based electronic nano-devices. For high qulity electronic devices, the Joule heating induced breakdown is a significant bottleneck to the advancement of the devices,\cite{JaveyA2004,PopE2005,PopE2007,BaeMH,Ong} so it is of essential importance to enhance the capability of heat delivery in these devices. The situation becomes completely opposite in the thermoelectric field, where low thermal conductivity is in great demand for high value of ZT.\cite{NolasGS2001} Hence, it is an urgent and practical task to supply some information for experimentalists on the modification of the heat transport capability for their graphene and hexagonal boron nitride (CBN) compound for different application motivations. That is the aim of present work.

In this work, we use both molecular dynamics (MD) simulation and nonequilibrium Green's function (NEGF) approaches to investigate the heat transport across the CBN interface. The interface is built inner a graphene nanoribbon of armchair edge. We study the heat current reduced by CBN interface of different shapes and sizes, including zigzag trigonal interface [CBN(TZ)], armchair trigonal interface [CBN(TA)], square interface [CBN(S)], and circular interface [CBN(C)]. The size of the interface is characterized by the number of carbon atoms at the interface, $N_{\rm edge}$, which varies from 0 to 60 in this work. Two same phenomena are observed from the two different investigation methods. Firstly, the heat current is reduced linearly with the increase of $N_{\rm edge}$ in all samples, and the reduction is more than 50\% for $N_{\rm edge}\approx 50$. Secondly, with same $N_{\rm edge}$, the zigzag interface causes stronger reduction of heat current than the armchair interface. These effects are interpreted from the lattice dynamics analysis and the NEGF transmission function which are nicely consistent with each other. The lattice dynamics analysis finds that the heat reduction is induced by the localized phonon modes at the CBN interface. The density of states (DOS) for the localized modes increases with increasing $N_{\rm edge}$ and it is larger in zigzag interface than the armchair interface, which explain the two observed phenomena. Particularly, strong localized modes are found in the frequency regions around 300, 800, and 1400 cm$^{-1}$. In the other explanation, the NEGF transmission function is depressed more seriously with increasing $N_{\rm edge}$, and stronger depression is found in the zigzag interface than the armchair interface. It also shows that the transmission function of phonon modes around 300, 800, and 1400 cm$^{-1}$ undergo strong depression for all CBN interfaces, which coincides with the former lattice dynamics analysis.

\begin{figure}[htpb]
  \begin{center}
    \scalebox{1.0}[1.0]{\includegraphics[width=8cm]{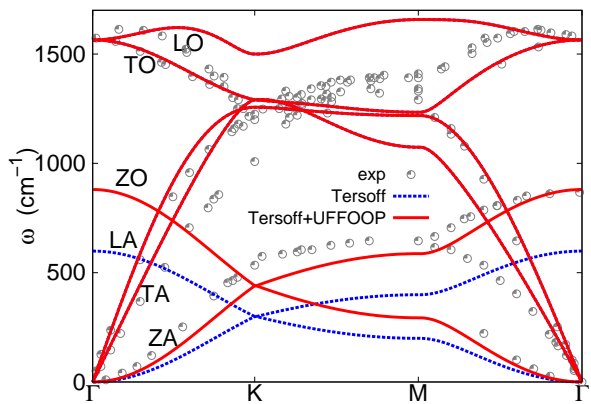}}
  \end{center}
  \caption{(Color online) Phonon spectrum of graphene calculated from Tersoff potential (dotted blue line) and Tersoff+UOOP potential (solid red line). Experimental data are read from figure 2 of Ref.~\onlinecite{Mohr}}
  \label{fig_graphene_phonon}
\end{figure}
\section{interatomic potential}
There are three different elements, C, B, and N, in present work. We have recently developed a set of Tersoff+UOOP potential to simultaneously describe both the in-plane and out-of-plane vibrations in hexagonal boron nitride sheet.\cite{JiangJW2011} This potential is of high efficiency, stability, and is particularly suitable for the investigation of heat transport as it properly describes the phonon spectrum. The interaction between B and N in the CBN samples is governed by this set of Tersoff+UOOP potential. For the interaction between carbon atoms in graphene, we can develop a parallel set of Tersoff+UOOP potential as the graphene and BN sheet have quite similar honeycomb structure. The original Tersoff parameters for carbon system were developed for sp3 structures,\cite{Tersoff} so they can not be directly applied to sp2-bonding graphene. Explicitly, the frequencies of in-plane vibrations from this potential are about 60\% higher than experimental values. To eliminate this flaw, we scale the $A$ and $B$ parameters in Tersoff potential by a factor of 0.4, which leads to reasonable frequencies for in-plane vibrations as shown in Fig.~\ref{fig_graphene_phonon} by dotted line (blue online). However, the result also shows that the frequencies of out-of-plane vibrations (ZA and ZO) are obviously smaller than the experimental value. This disagreement roots in the form of the Tersoff potential, where the in-plane and out-of-plane vibrations are coupled together. To modify the out-of-plane vibrations without affecting the in-plane vibrations, we introduce the universal out-of-plane (UOOP) potential\cite{Gale} to describe the out-of-plane vibrations. The UOOP potential has the form $V=C_{0}(C_{1} + C_{2}\cos(\phi) + C_{3}\cos(2\phi))$, where $\phi$ is the dihedral angle. The four optimized parameters $C_{i}$ are $-$0.588047 eV, 1.000000, 0.814936, and 0.259776, respectively. This set of Tersoff+UOOP potential results in good phonon spectrum of graphene as shown in Fig.~\ref{fig_graphene_phonon} by solid red line, so it is suitable for the study of heat transport in graphene. Note, however, that the calculated frequencies of the high energy in-plane optical modes (LO and TO) do not agree with experiment in detail, because these two optical modes are more sensitive to long-range interaction which is ignored in the bond order Tersoff potential. The interactions between different elements are also described by the Tersoff potential with the parameters determined by the combining law of multicomponent systems.\cite{Tersoff1989}
\begin{figure}[htpb]
  \begin{center}
    \scalebox{1.0}[1.0]{\includegraphics[width=8cm]{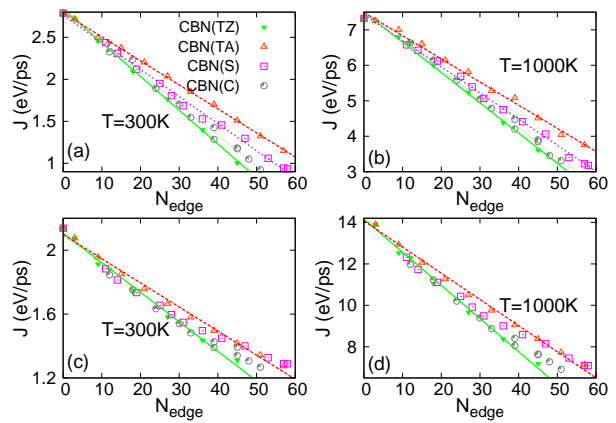}}
  \end{center}
  \caption{(Color online) The heat current v.s $N_{\rm edge}$ for different CBN interfaces: (a) and (b) from MD simulation; (c) and (d) from NEGF calculation.}
  \label{fig_J}
\end{figure}
 The single parameter $\chi$ introduced by the combining law is set to be 1.0, which may lead to possible error. However, the error is quite small, because $\chi$ acts as a fine tuning parameter. Furthermore, the success of the combining law mainly relies on the Tersoff potentials between individual element, which have been carefully fitted to experimental data in the above.

\section{results and discussion}
We run MD simulation to calculate the heat current $J$ across the CBN compounds. $J$ is calculate from $J=(J_{L}-J_{R})/2$, where $J_{L/R}$ are the heat energy pumped into system from left/right heat baths. The temperatures of two baths are $T_{L/R}=(1\pm \alpha)T$ with $\alpha=0.1$. The constant temperature of heat bath is realized by the No\'se Hoover thermostat.\cite{Nose,Hoover} A time step of 1.0 fs is used. The simulation is run at least for $2\times10^{8}$ MD steps and will be sufficiently extended to ensure that the steady state is achieved. The system is in steady state when the difference of heat current from left and right baths is less than 2\% of the current across the system. We study the heat current in the graphene nanoribbon with armchair edge. The length of the nanoribbon is 102.24~{\AA} and the width is 39.36~{\AA}. Some C atoms in the center region are substituted by B and N atoms. The BN regions in the graphene nanoribbon can be of different shapes and sizes. Periodic boundary condition is applied in the lateral direction of the nanoribbon, so that we can avoid the edge effect and focus on the effect of the CBN interface. We have also performed calculation for graphene nanoribbon with zigzag edge and the same size, where similar phenomenon was observed.
\begin{figure}[htpb]
  \begin{center}
    \scalebox{1.0}[1.0]{\includegraphics[width=8cm]{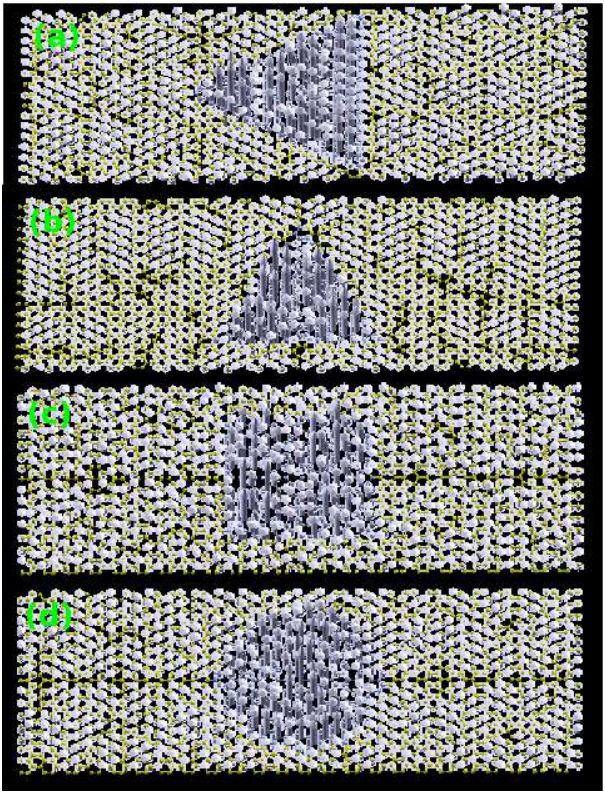}}
  \end{center}
  \caption{(Color online) Vibrational morphology of some selected localized phonon modes in different CBN compounds with $N_{\rm edge}=45$. The CBN compounds from (a)-(d) are CBN(TZ), CBN(TA), CBN(S), and CBN(C).}
  \label{fig_u}
\end{figure}
 Fig.~\ref{fig_J}~(a) and (b) show the heat current $J$ across four CBN interface of various sizes. With the increase of $N_{\rm edge}$, a linear decrease of $J$ is observed for all CBN interfaces. The MD simulation results are displayed by points, and the lines are the linear fitting of these MD results. For $N_{\rm edge}\approx 50$, the heat current $J$ is reduced by more than 50\%. With same value of $N_{\rm edge}$, the CBN(TZ) interface causes the strongest reduction to heat current, while the CBN(TA) interface introduces the weakest reduction. The heat currents across the CBN(S) and CBN(C) interfaces are in the middle of these two trigonal samples, where the reduction of $J$ by CBN(C) interface is slightly stronger than CBN(S) interface. All these phenomena are the same at both 300 K and 1000 K, except different value for $J$. The difference between heat current in zigzag and armchair types of CBN interfaces increases with increasing $N_{\rm edge}$. A quite large difference is desirable in experimental CBN compounds as $N_{\rm edge}$ is usually very large.\cite{CiL} \textbf{We note that the quantity $N_{edge}$ is an appropriate choice to distinguish the cluster types. If the thermal current is plotted as a function of cluster area, it is almost a single valued function of the cluster area, which implies that the cluster area also plays a role besides the number of edge atoms.}

\begin{figure}[htpb]
  \begin{center}
    \scalebox{1.0}[1.0]{\includegraphics[width=8cm]{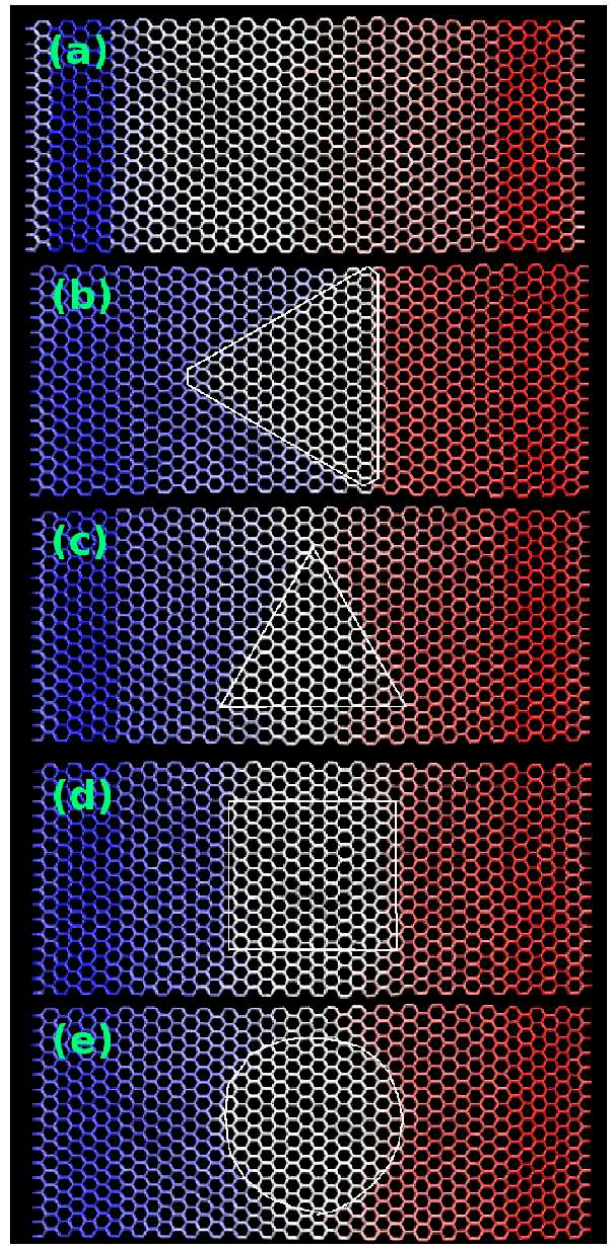}}
  \end{center}
  \caption{(Color online) Temperature distribution for different CBN compounds: (a). pure graphene; (b). CBN(TZ); (c). CBN(TA); (d). CBN(S); (e). CBN(C). Temperature decreases from 330 K to 270 K from blue to red.}
  \label{fig_dTdx}
\end{figure}
\begin{figure}[htpb]
  \begin{center}
    \scalebox{1.0}[1.0]{\includegraphics[width=8cm]{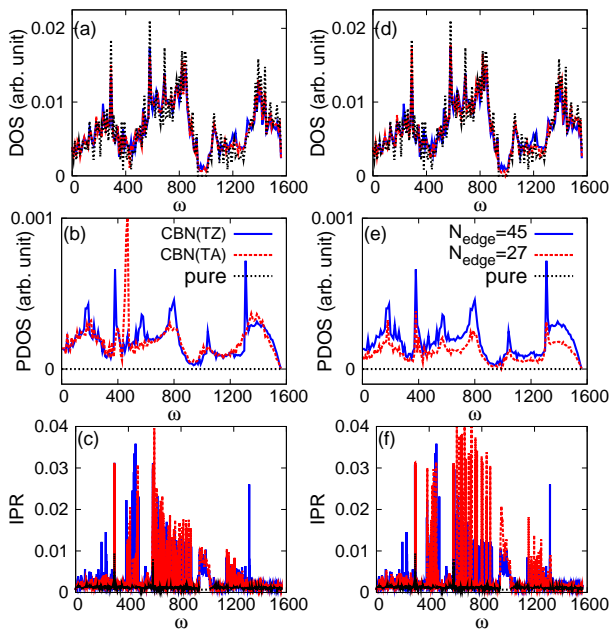}}
  \end{center}
  \caption{(Color online) Localization properties for phonon modes in different CBN interfaces. (a)-(c) are the total DOS, projected DOS, and IPR of CBN(TZ), CBN(TA), and pure graphene. (d)-(f) are for CBN(TZ) with $N_{\rm edge}$=45, 27, and pure graphene.}
  \label{fig_dos}
\end{figure}
A straight forward physical interpretation for the heat reduction can be made from the lattice dynamics analysis. There are some localized modes around the CBN interface. These modes do not move and make few contribution to heat transport. Fig.~\ref{fig_u} show selected localized modes for all studied CBN interfaces. In these modes, only boron and nitride atoms vibrate while all carbon atoms do not take part in the vibration. As a result of these localized modes, there will be temperature jumps at the CBN interfaces as shown in Fig.~\ref{fig_dTdx}. For different types of CBN interfaces with the same $N_{\rm edge}$, the number of atoms per length at the interface is the largest (smallest) in zigzag (armchair) type of CBN interface. As a result, the DOS of the localized modes in the CBN(TZ) is larger than that of the CBN(TA), so the CBN(TZ) interface induces stronger heat reduction than CBN(TA). The CBN(S) and CBN(C) interfaces are a mixture of zigzag and armchair types; thus their reduction capability is in the middle between the two trigonal samples of pure zigzag or armchair interfaces. In Fig.~\ref{fig_dos}~(a), (b), and (c), we numerically compare the localized modes in CBN(TZ), CBN(TA), and pure samples. Panel (a) shows that all three samples have similar total DOS. Panel (b) is the DOS projected (PDOS) onto the CBN interface. There is no CBN interface in pure sample, so PDOS is zero in whole frequency range. The PDOS in CBN(TZ) is larger than the CBN(TA) due to larger number of atoms per length at the CBN(TZ) interface, which confirms our above qualitative argument. Panel (c) shows the inverse participation ratio (IPR) of each mode. The IPR value of the localized modes is about one or two orders larger than transmissible modes.\cite{BellRJ} There is no localized mode in pure sample, and the IPR in whole frequency range is very small. In the two trigonal samples, there are lots of localized modes. Especially, those localized modes in the frequency range of 300, 800, and 1400 cm$^{-1}$ have very high IPR. Panel (b) and (c) together manifest that the zigzag interface can introduce stronger heat reduction. For a particular sample, the length of CBN interface increases with increasing $N_{\rm edge}$. This leads to more localized modes at the CBN interface, resulting in stronger reduction. Fig.~\ref{fig_dos}~(d), (e), and (f) compare the DOS, PDOS and IPR in CBN(TZ) interface with $N_{\rm edge}=27$, 45, and pure samples. Panel (e) shows that the PDOS in the sample with $N_{\rm edge}=45$ is obviously larger
\begin{figure}[htpb]
  \begin{center}
    \scalebox{1.0}[1.0]{\includegraphics[width=8cm]{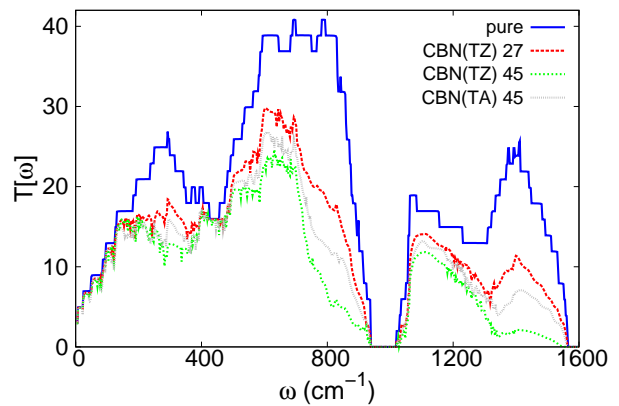}}
  \end{center}
  \caption{(Color online) The NEGF transmission function for CBN interface of different shapes and sizes.}
  \label{fig_transmission}
\end{figure}
 than that of $N_{\rm edge}=27$, resulting in much stronger reduction to $J$.

In the above, we have run MD to simulate the heat current adjusted by various CBN interfaces, and provided an explanation by means of the lattice dynamics analysis. However, there is no phonon-phonon scattering effect in the localized modes, while this effect is considered in MD simulation. To reinforce our conclusions from MD simulation and the explanations from localized modes, we repeat the whole calculation using the ballistic NEGF approach.\cite{WangJS2008} The ballistic NEGF do not take into account the phonon-phonon scattering. But it is an exact treatment for the scattering at the CBN interface, which actually reflects the localization property of the localized modes. If the NEGF leads to the same conclusions as MD simulation, then we can safely announce that the phonon-phonon scattering in MD simulation has no effect on heat reduction. Fig.~\ref{fig_J}~(c) and (d) show the heat current calculated from NEGF, which are consistent with MD simulation results. Firstly, $J$ decreases almost linearly with increasing $N_{\rm edge}$ for a particular sample. For $N_{\rm edge}\approx 50$, the heat current is reduced by more than 50\%. Secondly, for the same $N_{\rm edge}$, the CBN(TZ)/CBN(TA) interfaces cause the strongest/weakest reduction of heat current. Heat reduction from the other two interfaces are between these two trigonal interfaces. These two results are the same at both 300 K and 1000 K. It should be note that the value of $J$ from NEGF approach is almost twice of that from MD simulation at 1000 K, because of strong phonon-phonon scattering in MD simulation at high temperatures. The value of $J$ from two different methods is close to each other at 300 K, where the phonon-phonon scattering is weak. We have obtained the same conclusions from MD simulation and NEGF calculation, although the values of $J$ are quite different from these two approaches. We further show the transmission function in the whole frequency range from NEGF method in Fig.~\ref{fig_transmission}. It shows that the transmission function in CBN(TZ) is smaller than that
\begin{figure}[htpb]
  \begin{center}
    \scalebox{1.0}[1.0]{\includegraphics[width=8cm]{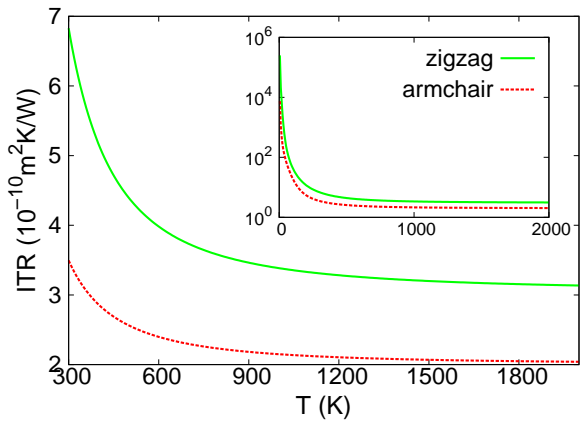}}
  \end{center}
  \caption{(Color online) The interface thermal resistance v.s temperature for zigzag (solid line, green online) and armchair (dotted line, red online) CBN interfaces. Inset shows the interface thermal resistance (in $log_{10}x$ scale) in whole temperature range.}
  \label{fig_itr}
\end{figure}
 of the CBN(TA) with same $N_{\rm edge}$. The transmission function is smaller in samples with larger $N_{\rm edge}$, indicating stronger heat reduction in these samples. A general feature is that the transmission functions for phonon modes with frequencies around 300, 800, and 1400 cm$^{-1}$ are depressed seriously by CBN interfaces. This feature agrees quite well with the strong localized modes in these frequency regions as we have discussed previously. These results further confirm that the reduction of heat current is because of the localized modes at the CBN interface. Fig.~\ref{fig_transmission} also shows that high frequency modes, which have increased contribution to heat current at higher temperatures, are suppressed more effectively. As a result, the reduction is much more effective at high temperatures in Fig.~\ref{fig_J}, as indicated by the larger slope of the fitted line at 1000 K than that of the 300 K.

The NEGF approach can provide us another more immediate evidence--interface thermal resistance (ITR)--for the different heat reduction in zigzag and armchair type CBN interfaces. We apply NEGF to study heat transport across junctions formed by graphene on left part and BN sheet on the right part, with the CBN interface at the middle. NEGF calculates the thermal conductance ($\sigma$) of the junction. The thermal resistance can be obtained through $R=s/\sigma$, where $s$ is the cross section of the junction considering 3.35~{\AA} as the thickness. The ITR can be estimated by ITR$=R-R_{0}$, with $R_{0}$ as the thermal resistance in the pure graphene sample. Fig.~\ref{fig_itr} compares the ITR for zigzag and armchair CBN interfaces. The ITR in zigzag CBN interface is about $7\times10^{-10}$m$^{2}$K/W at room temperature and decreases to $3.5\times10^{-10}$m$^{2}$K/W at higher temperatures. The value of ITR in armchair CBN interface is only about half of the zigzag CBN interface in whole temperature range. As a direct result from its higher ITR, the zigzag CBN interface can introduce stronger heat reduction than the armchair CBN interface.

\section{conclusion}
To conclude, this work studies how the heat current is manipulated by the CBN interface of various shapes and sizes. The interatomic interaction is described by a set of Tersoff+UOOP potential, which is efficient, stable, and suitable for the study of heat transport in the sp2-bonding system. The calculation is accomplished by using two distinct approaches: MD simulation and NEGF calculation. Both methods find same phenomena that the heat current is reduced linearly with the increase of $N_{\rm edge}$, and the reduction can be more than 50\% for $N_{\rm edge}\approx 50$. It is also shown that the zigzag CBN interface introduces stronger reduction to heat current than the armchair CBN interface. These observations are explained from two different points of view, which are eventually consistent with each other. Both explanations interpret that the localized phonon modes at the CBN interface play a key role in the heat reduction and the phonon-phonon scattering has no effect. A more immediate evidence from NEGF for the stronger heat reduction in zigzag CBN interface is its stronger ITR, which is nearly double of that in the armchair CBN interface.

The graphene and BN sheet are two promising nano-materials with plenty of advance properties. They have similar honeycomb structures with close configuration parameters.\cite{Dubay,Kern} Existing works have demonstrated that the compounds of these two materials can upgrade the electronic properties in graphene and particularly tunable electronic band gap can be realized. The practical significance of present work is to provide a guideline for the experimentalists to realize tunable heat transport capabilities through shaping the CBN interfaces during the synthesis of CBN compounds. So that higher heat transport capability can be achieved for rapid electronic devices and lower heat transport capability is obtained for thermoelectric nano-devices.

\textbf{Acknowledgements} The work is supported by a URC grant of R-144-000-257-112 of National University of Singapore.

\end{document}